\documentclass[preprint,12pt]{elsarticle}

\usepackage{amssymb}
\usepackage{amsmath}
\usepackage{float}
\usepackage{xcolor}

\usepackage{siunitx}
\usepackage{tikz}

\usepackage{lineno}

\usepackage{subcaption}
\usepackage{pdfpages}

\journal{Nuclear Instruments and Methods}

\begin{document}

\begin{frontmatter}





\title{Diamond-like carbon coatings for cryogenic operation of particle detectors} 

\author[1]{S. Leardini\fnref{fnref1}}
\author[3,4]{Y. Zhou}
\author[2]{A. Tesi}
\author[1]{M. Morales}
\author[1]{D. González-Díaz}
\author[2]{A. Breskin}
\author[2]{S. Bressler}
\author[2]{L. Moleri}
\author[1]{V. Peskov}

\address[1]{IGFAE, Universidade de Santiago de Compostela, ES} 
\address[2]{Weizmann Institute of Science, Department of Particle Physics and Astrophysics, Rehovot, IL} 

\address[3]{State Key Laboratory of Particle Detection and Electronics, University of Science and Technology of China, Hefei 230026, China}
\address[4]{Department of Modern Physics, University of Science and Technology of China, Hefei 230026, China}
\fntext[fnref1]{Sara.Leardini@usc.es}

\begin{abstract}

Characterization of diamond-like carbon (DLC) coatings at cryogenic temperatures (down to 77~K) is presented, covering the electrical resistivity range of practical interest to gaseous and liquid particle instrumentation: $10^{-1}$-$10^5$~M$\Omega$/$\Box$. The good behaviour observed in terms of linearity, surface uniformity and stability with time and transported charge add to other well-known characteristics like low chemical reactivity and tolerance to radiation. The observed temperature dependence and stability of electrical properties with transported charge is consistent with a conductivity mechanism based on 2-dimensional variable-range electron hopping, as expected for the surface conductivity of thin films made from amorphous carbon. First results from a resistive-protected WELL  detector ('RWELL') built with DLC and operated close to the liquid-vapor coexistence point of argon (87.5~K at 1~bar) are presented.

\end{abstract}


\end{frontmatter}

\section{Introduction}

\begin{figure}[H]
    \centering
    \includegraphics[width=15cm]{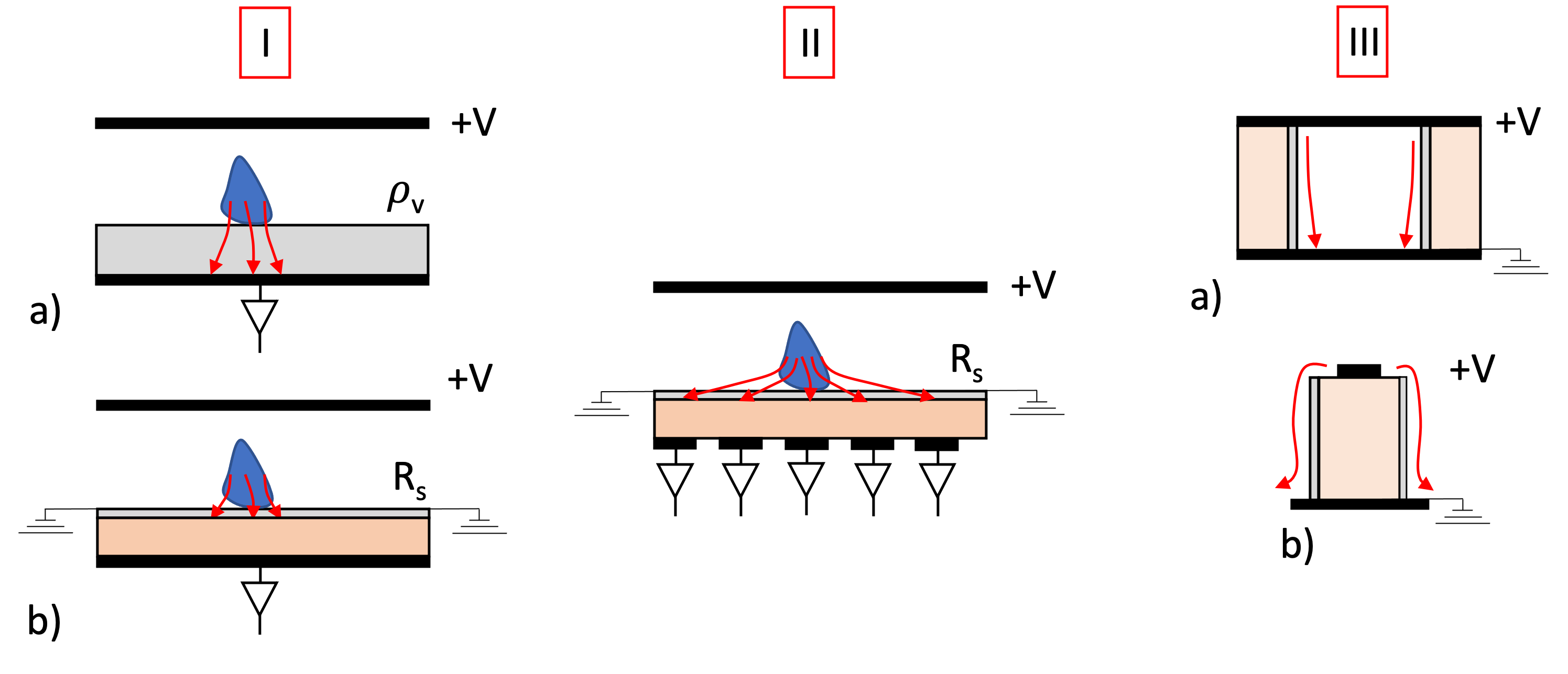}
     
    \centering
    \caption{\footnotesize Most common situations of practical interest for use of resistive materials in gaseous and liquid particle detection technology. Ia-Ib) `spark quenching': current flow is limited to a small area, reducing the electrostatic energy available in case an avalanche reaches a too large size. II) `cluster-size tuning': current flow is adjusted in order to facilitate signal induction over a larger region, leading to oversampling and better spatial reconstruction of the charge centroid. III) `mitigation of charging-up and surface discharges': current flow through uniform resistive materials minimizes charging-up and provides suitable boundary conditions matching the electrostatic potential in the active medium.} 
    \label{fig:CasesR}
\end{figure}

Resistive materials can enhance the operational characteristics of particle physics instrumentation operating under high electric fields, however rugged resistive materials with stable electrical properties are costly and scarce. In radiation-detector science, there seem to exist at least three situations of recurring technological interest: i) stabilization of the gas-discharge process at high electric fields (`spark-quenching'), \cite{Pestov, Santonico, Fonte, uRWELL}; ii) adjusting the charge-induction profile from moving charges to improve space resolution in tracking detectors \cite{Dixit, Riegler, T2K}; iii) reducing local charging-up as well as suppressing surface discharges in, for instance, high voltage feedthroughs \cite{sphere}, some types of micropattern gas detectors (MPGD) \cite{MSGCs, GEM-DLC}, and field-cages in time projection chambers \cite{DUNE-FC} (Fig. \ref{fig:CasesR}). While the response of a system in the above situations is dependent on the overall geometry, the main performance metric is usually the electrical resistivity of the material, either bulk $\rho_v$ or surface $R_S$. For linear, homogeneous and isotropic coatings, sheets or plates, they relate through $R_S = \rho_v / t$, where $t$ is the material thickness. Approximate orders of magnitude can be found in table \ref{tab:data}, for either coatings (`films') or plates. With a `too-conductive' material potentially leading to discharges, intolerable currents or the impossibility of localizing a moving charge (depending on the application), a `too-resistive' material would lead, on the other hand, to surface charging-up and field deformations. It thus follows that materials of interest to particle-physics instrumentation fall into the category of `bad insulators' (e.g., \cite{DGD_rev} and references therein). Because of the low industrial interest on this type of materials their availability is limited, calling for new materials and techniques for scientific use. An example of such focused R\&D effort for resistive plate chambers (RPCs) operated at room temperature is summarized in \cite{DGD_CBM}, however only limited effort has been made so far for resistive detectors operating under cryogenic conditions (e.g., \cite{Roy} for LXe temperature).

Diamond-like carbon (DLC) coatings represent a powerful technology, that has been historically used for instance at mitigating surface discharges in the first MPGD detectors (micro-strip gas chambers or MSGC \cite{MSGCs}), and more recently have shown to reduce the transient gain behaviour characteristic of conventional FR4-based 'thick' gas electron multipliers (THGEMs), \cite{GEM-DLC}. They have been used for resistive-protected MPGD detectors as well, for instance in $\mu$-RWELL configuration \cite{uRWELL}, and in resistive plate chambers \cite{RPC-DLC} or as robust photo-cathodes \cite{Piantanida,picosec}. 

One problem with DLC coatings, as with most insulating materials, is that their resistivity quickly rises at cryogenic temperatures: a material suitable for operation at room temperature can become a perfect insulator in cryogenic conditions, impairing operation. Thus in this work we consider thin DLC films made by sputtering on Kapton and FR4 substrates, and study their behaviour in the surface resistivity range $0.1$-$10^5$ \textnormal{M}$\Omega$/$\Box$, down to liquid nitrogen ($LN$) temperature. The chosen range is expected to cover most situations of practical interest. Uniformity, stability with time and transported charge, and dependence with applied voltage are discussed. Validation results from a DLC-based RWELL detector operated at near LAr temperature are presented.

\begin{table*}
\setlength{\tabcolsep}{11pt}
\renewcommand{\arraystretch}{1.3}
\centering
\begin{tabular}{c| c| c}
\hline \hline
$\textnormal{case}$ & $R_S [\textnormal{M}\Omega/\Box]$ & $\rho_v [\textnormal{G}\Omega~\cdot\textnormal{cm}]$ \\
\hline
$Ia$   &	-                &	$0.1$-$1$ to $10^3$-$10^4$\cite{Francke}  \\
$Ib$   &	$10$-$1000$ \cite{GianniSystematic} &	 -             \\
$II$   &	$0.5$-$2.5$ \cite{Dixit,T2K}       &	- \\
$IIIa$ &	$1000$-$10000$ \cite{DUNE-FC, Achiles}          &	- \\
$IIIb$ &	$1000$-$10000$ \cite{sphere}               &	- \\
\hline
\hline
\end{tabular}
\caption{\label{tab:data} Surface and bulk resistivity values employed by various authors in most typical applications. In I) the lower limit is imposed by loss of spark quenching and the upper one by loss of rate capability (charge build-up). For typical geometries, the values given in II) lead to cluster sizes at the mm's scale. In III) the lower limit is given by current consumption, the upper limit is setup-dependent.}
\end{table*}

\section{Experimental methods}\label{sec:setup}

\subsection{DLC coating}

\begin{figure}[H]
    \centering
    \includegraphics[width=15cm]{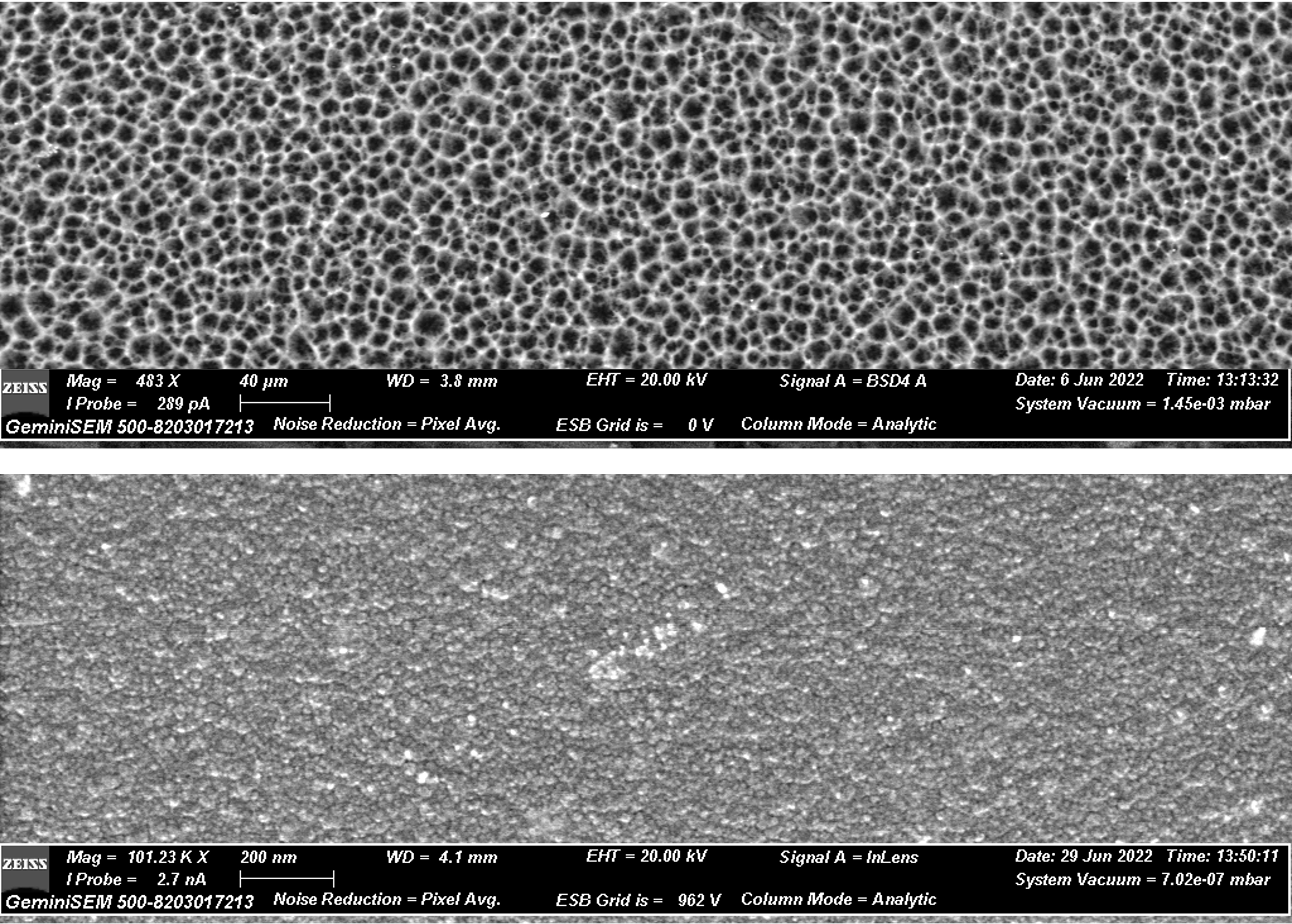}
    \centering
    \caption{\footnotesize Images taken with scanning electron microscope of the DLC films on FR4 ($40\mu$m scale, top) and Kapton ($200nm$ scale, bottom). The scales have been chosen to highlight the main features observed.}
    \label{fig:SEM}
\end{figure}

DLC was deposited through magnetron sputtering \cite{GEM-DLC}. A Teer 650 device (from Teer Coating Ltd) at the SKL in Hefei was used, allowing evaporations up to surface dimensions of 25 × 25~cm$^2$. The substrates were cleaned with ethanol and dried in an oven at 70 $^\circ$C. The sputtering lasted for about 20-60 min, resulting in film thicknesses generally in the range 10's-100's nm (more details on the process can be found on \cite{GEM-DLC}). Evaporations were made on FR4 and Kapton, two common detector substrates of different mechanical and surface properties (Fig. \ref{fig:SEM}). 
In total, six DLC-coated Kapton samples of different surface resistivities were produced, named with letters $A$-$F$, and two DLC-coated FR4 samples, $a$ and $f$, made with sputtering conditions similar to $A$ and $F$. Different film thicknesses and resistivities were obtained by changing the sputtering time.
As a reference, the thickness of the most conductive FR4 sample presented in this work was estimated to be compatible with $t=160\pm10$~nm, the uncertainty reflecting the variation between the in-situ measurement (with a step profiler) and the ex-situ one (by grazing-angle X-ray reflectometry on a twin sample evaporated on silicon substrate).

\subsection{Cryogenic setup}

\begin{figure}[H]
    \centering
    \includegraphics[width=12.5cm]{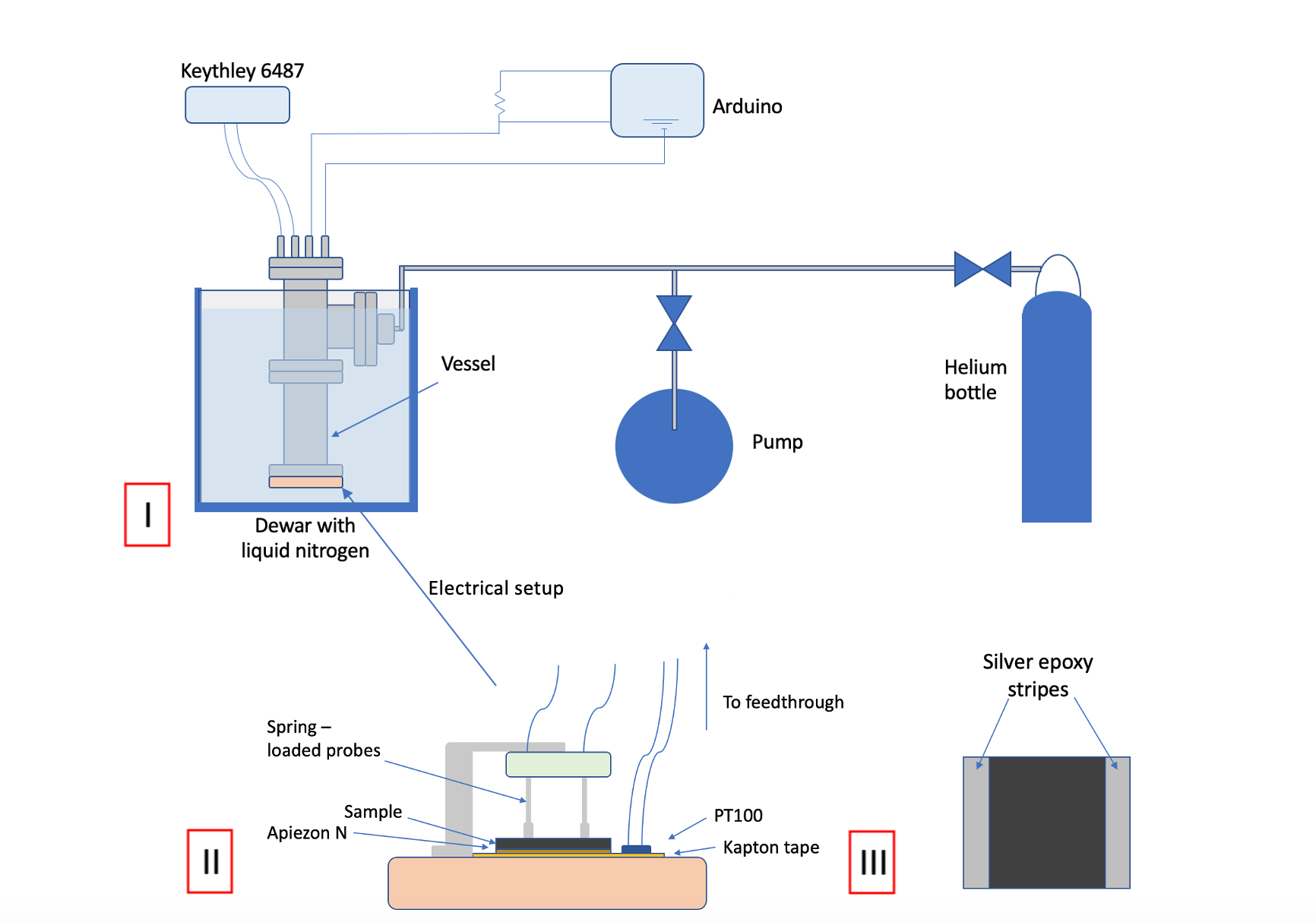}
     
    \centering
    \caption{\footnotesize Scheme of the setup used for electrically characterizing DLC-coated samples. I) Full view: the sample is laying on the part of the vessel colored in orange. The vessel is immersed in liquid nitrogen inside a Dewar, and connected to a Pfeiffer Duo 016B pump and a helium bottle. It also communicates with a Keythley 6487 and an Arduino thanks to the upper feedthrough. II) Close-up of the electrical setup inserted into the vessel: a Kapton layer electrically insulates the sample from the vessel bottom, while an additional layer of Apiezon N grease improves the thermal coupling. The sample is connected to two spring-loaded probes to measure the surface resistivity; close to the sample, a PT100 is used to measure the temperature inside the vessel. III) Top view of the sample: the two silver epoxy electrodes define a 1 cm $\times$ 1 cm square.}
    \label{fig:setup1}
\end{figure}

The surface resistivity as a function of temperature was determined with a physical property measuring system (PPMS, Model 6000) from Quantum Design, following a square four-point probe method over a $\sim 1~$cm$^2$ area. Geometrical corrections were applied according to \cite{Miccoli2015}, amounting to about 30\%. The PPMS assures thermal equilibrium at every measuring step, however its sensitivity reaches just the lower-end of resistivities of interest (up to $R_S = 30$ \textnormal{M}$\Omega$/$\Box$), so a system was devised to cover the entire range. Measurements were made in this case through biasing a silver epoxy strip on the sample and reading the return current on a second one, approximately forming a $1$~cm$^2$-square, through a Keithley 6487 picoammeter. The setup was assembled on a stainless steel vessel filled with a helium atmosphere (1.5~bar) and immersed in a $LN$-filled Dewar acting as a cryostat. To mitigate the effect of water freezing, the vessel was evacuated down to $10^{-3}$~mbar with a Pfeiffer Duo 016B pump prior to filling. The setup consisted of two spring-loaded probes in a PEEK frame, with a holding arm bolted to the vessel. The sample laid on the floor, electrically insulated with a thin Kapton layer and covered with APIEZON N grease to ease thermal coupling between sample and vessel. The temperature was sensed directly on the Kapton with a PT100 sensor, its resistance measured through an Arduino micro-controller board. A sketch of the setup is shown in Fig. \ref{fig:setup1}. 

\begin{figure}[H]
    \centering
     \includegraphics[width=12cm]{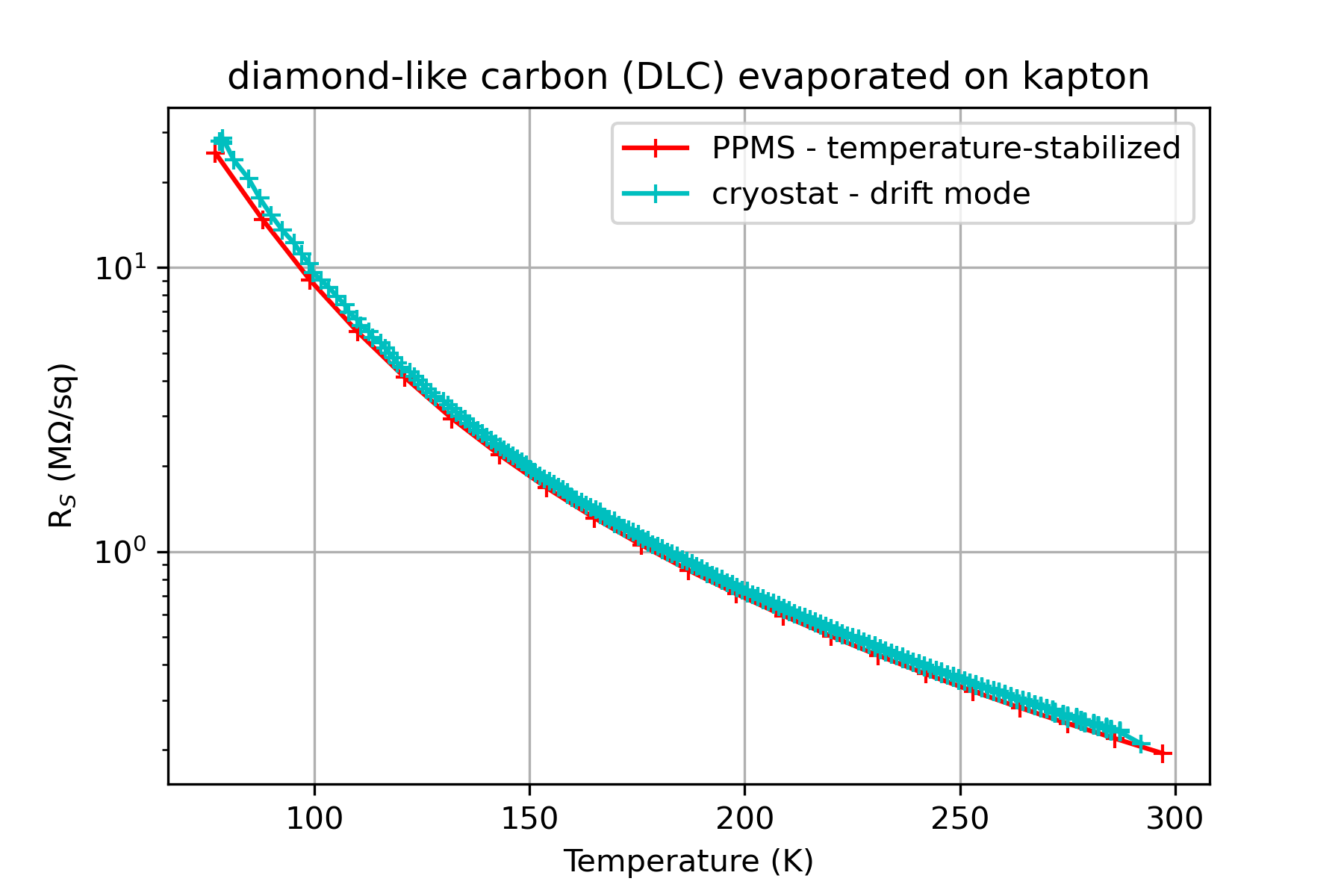}
    
\centering
    \caption{\footnotesize R-T curves for sample A as measured with PPMS (red) and with the dedicated setup described in section \ref{sec:setup}, developed for high-resistivity measurements (light blue).}
    \label{fig:clk_comm}
\end{figure}

The experimental procedure was the following: the vessel was immersed into $LN$ until $T$ = 77~K was measured in the PT100. At that point it was taken from the bath and let warm up to room temperature through its thermal inertia (`drift' mode). A comparison with the PPMS results for the least resistive sample showed an agreement within 10 \%, confirming that a sufficiently good thermal equilibrium is reached in the DLC samples through this procedure (Fig. \ref{fig:clk_comm}). Identical conclusions were extracted in a special PPMS run performed in drift mode, where the sample is also let warm up in a helium atmosphere, unimpeded. In the remainder of the text, error bars have been extracted from the standard deviation of two independent measurements.\footnote{This deviation must be understood as an estimate of the systematic uncertainty from the setup. A statistical interpretation in terms of confidence intervals is not straightforward without accomplishing a relatively high number of measurements, and was not dimmed necessary given the scope of the paper and the small relative deviations between measurements, generally at the level of 10-20\%.}

\section{Characterization of the samples \label{sec:homogeneity}}

\subsection{$I$-$V$ curves}
\begin{figure}[H]
    \centering
    \includegraphics[width=0.49\textwidth,]{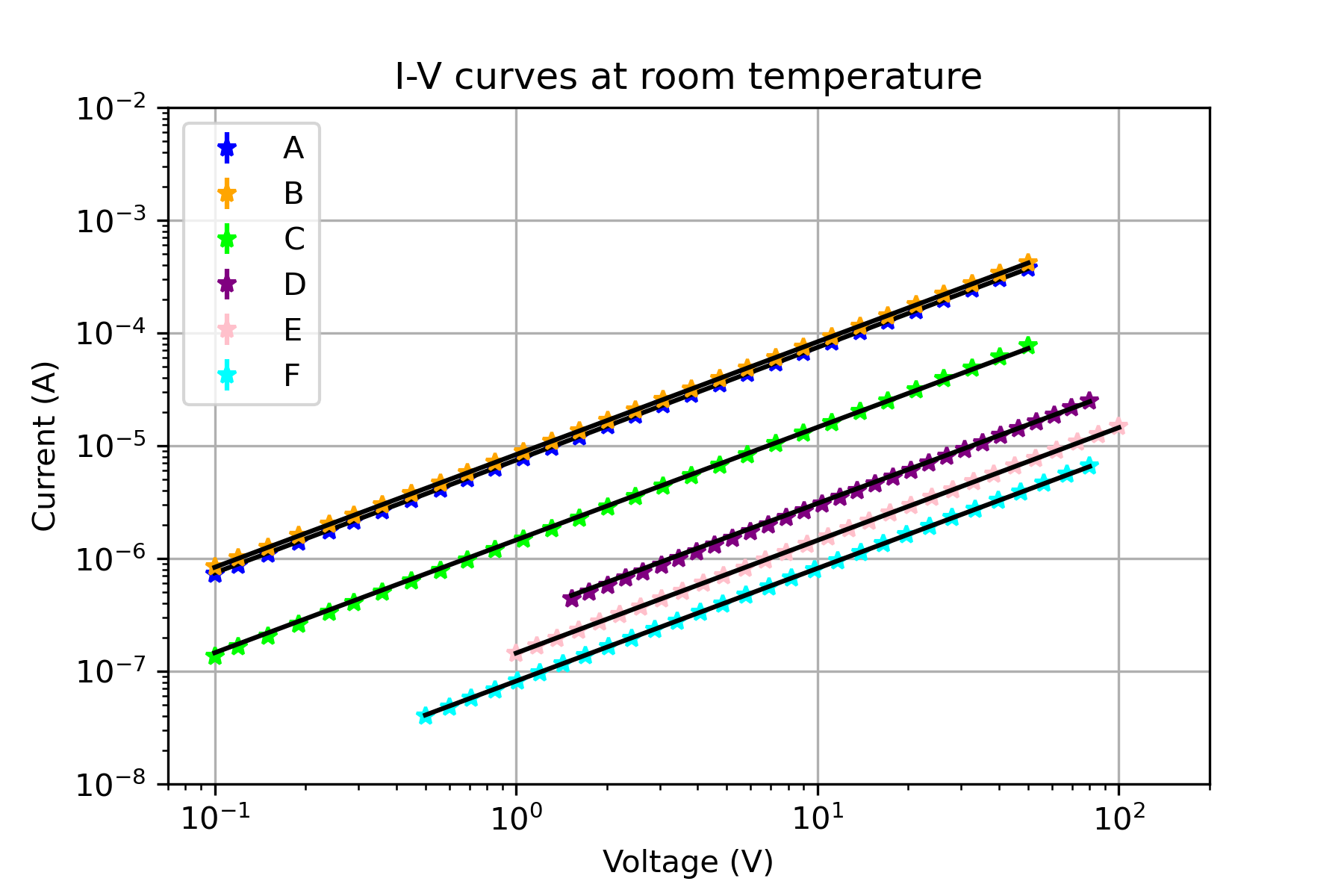}
    \includegraphics[width=0.49\textwidth]{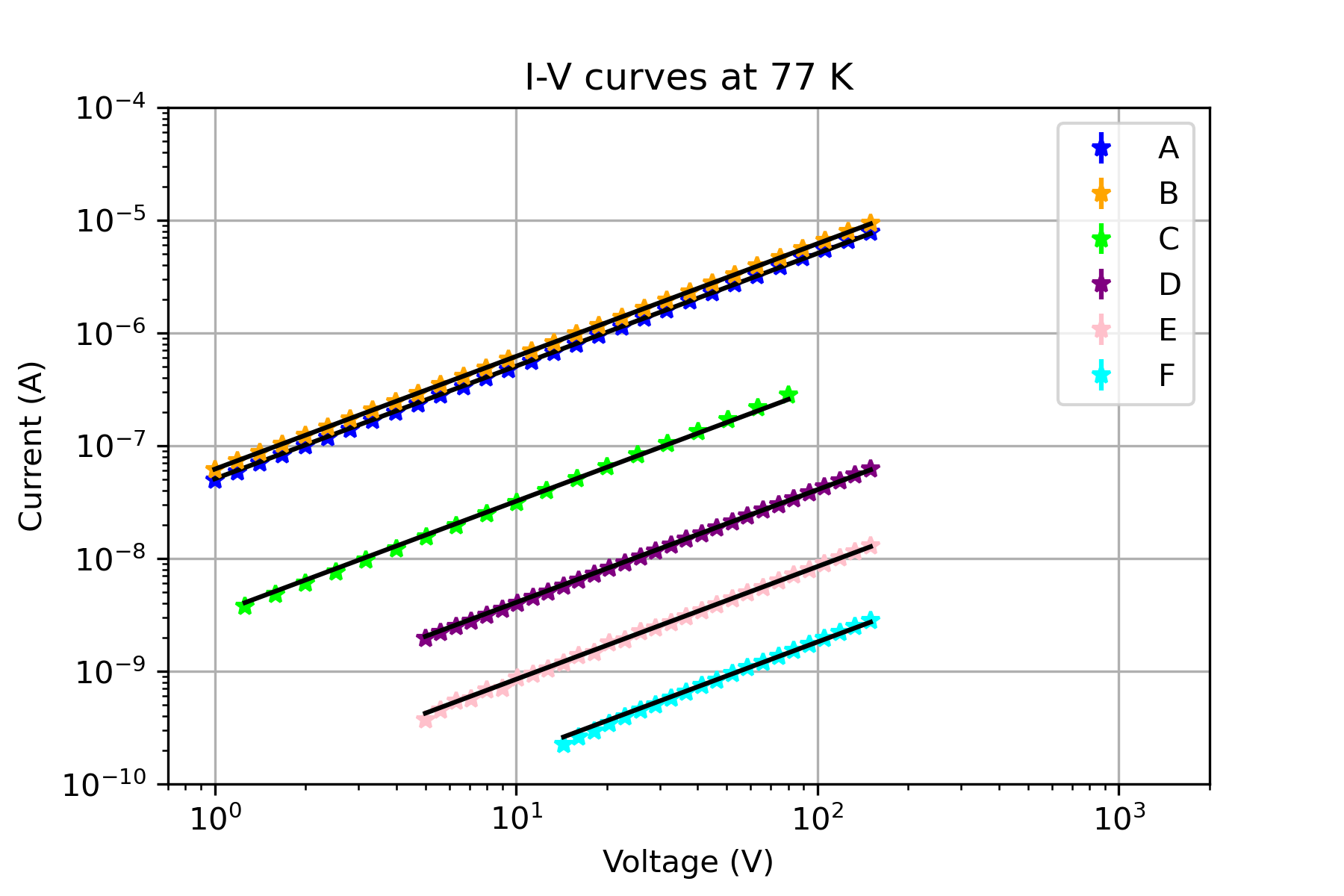}
    \caption{\footnotesize $I$-$V$ curves of the DLC-coated Kapton samples at room temperature (left) and liquid nitrogen temperature (right). Results for the FR4 samples ($a$, $f$) are not shown, but display the same trend. Super-imposed, the fit to $I = V/R$ lines.}
    \label{fig:linearity}
\end{figure}

The characteristic $I$-$V$ curves of the samples were obtained both at room and liquid nitrogen temperature, showing a perfect ohmic trend (Fig. \ref{fig:linearity}-lines represent the fit results). This simplifies the characterization as a function of temperature: the applied voltage was generally chosen to provide a good precision for the resistivity measurement at $LN$ temperature (starting point of the scan).

\subsection{Uniformity and dependence from substrate}
\begin{figure}[H]
    \centering
    \includegraphics[width=12cm]{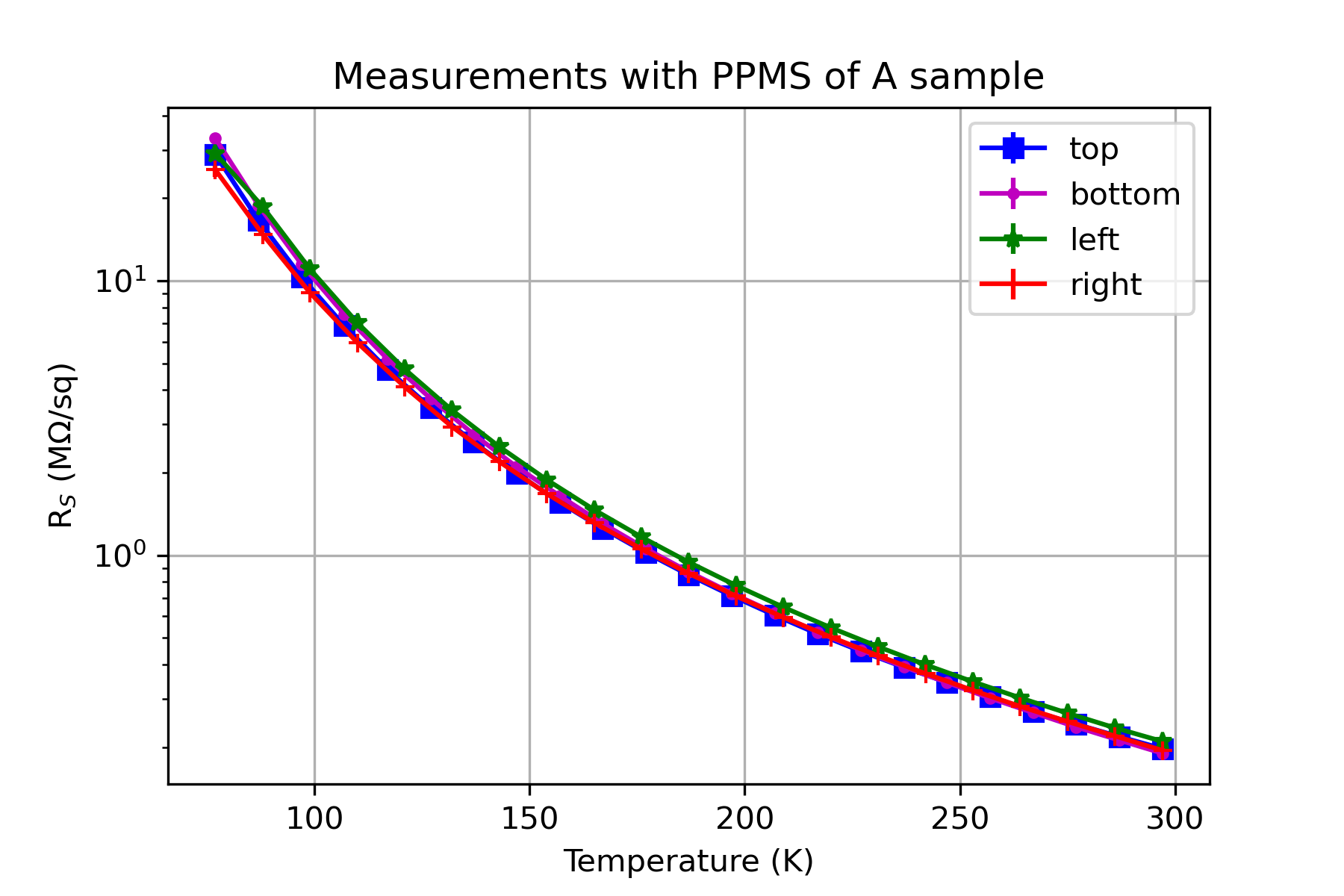}
   
    \centering
    \caption{\footnotesize $R$-$T$ curves from sample A taken with the PPMS device at the Magnetosusceptibility facility of USC, for 1~cm$^2$ regions chosen over the sample ($10 \times 10~$cm$^2$ in size).}
    \label{fig:homogeneity1}
\end{figure}

The electrical uniformity of the coatings was studied by measuring on four representative 1~cm$^2$-regions. Illustratively, results for sample A, 10 $\times$ 10~cm$^2$ in size, are shown in Fig. \ref{fig:homogeneity1}. The four measurements show an agreement within 20\%, down to $LN$ temperature, a result compatible with the 23\% uniformity values reported earlier in \cite{GEM-DLC}. The other samples were tested for uniformity at room temperature, providing analogous results.

\bigskip

\begin{figure}[H]
    \centering
    \includegraphics[width=12cm]{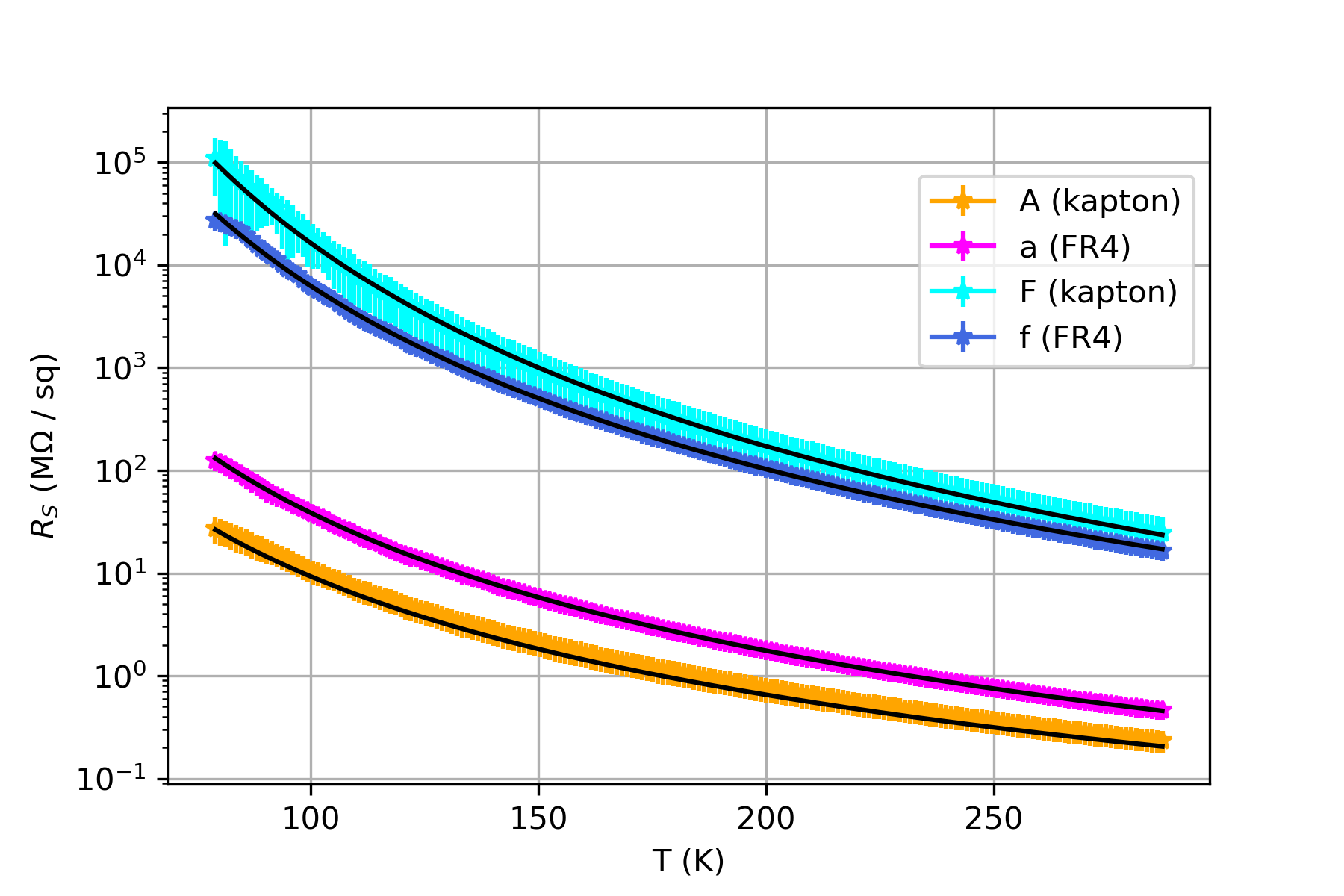}
\centering
    \caption{\footnotesize Comparison of samples with Kapton and FR4 substrates, grouped in couples for which similar parameters of the sputtering machine were attempted (A-a, F-f). The lack of a systematic deviation points to small changes of the sputtering process as the main source of discrepancy. The line represents the fit to the function described later in text.}
    \label{fig:batch}
\end{figure}

Kapton and FR4 are used regularly for manufacturing gaseous detectors, so they are interesting for understanding which role the substrate may play in the properties of the film. Fig. \ref{fig:batch} shows the result of such study for the couples A-a, F-f, made under similar sputtering conditions. Although the results are not compatible within the errors, the lack of a systematic deviation between FR4 and Kapton points to a small dependence with the substrate. Variations are likely due to small changes in the initial conditions of the sputtering machine.

\bigskip\noindent

\subsection{R-T curves \label{sec:resistivity}}
\begin{figure}[H]
    \centering
    \includegraphics[width=14cm]{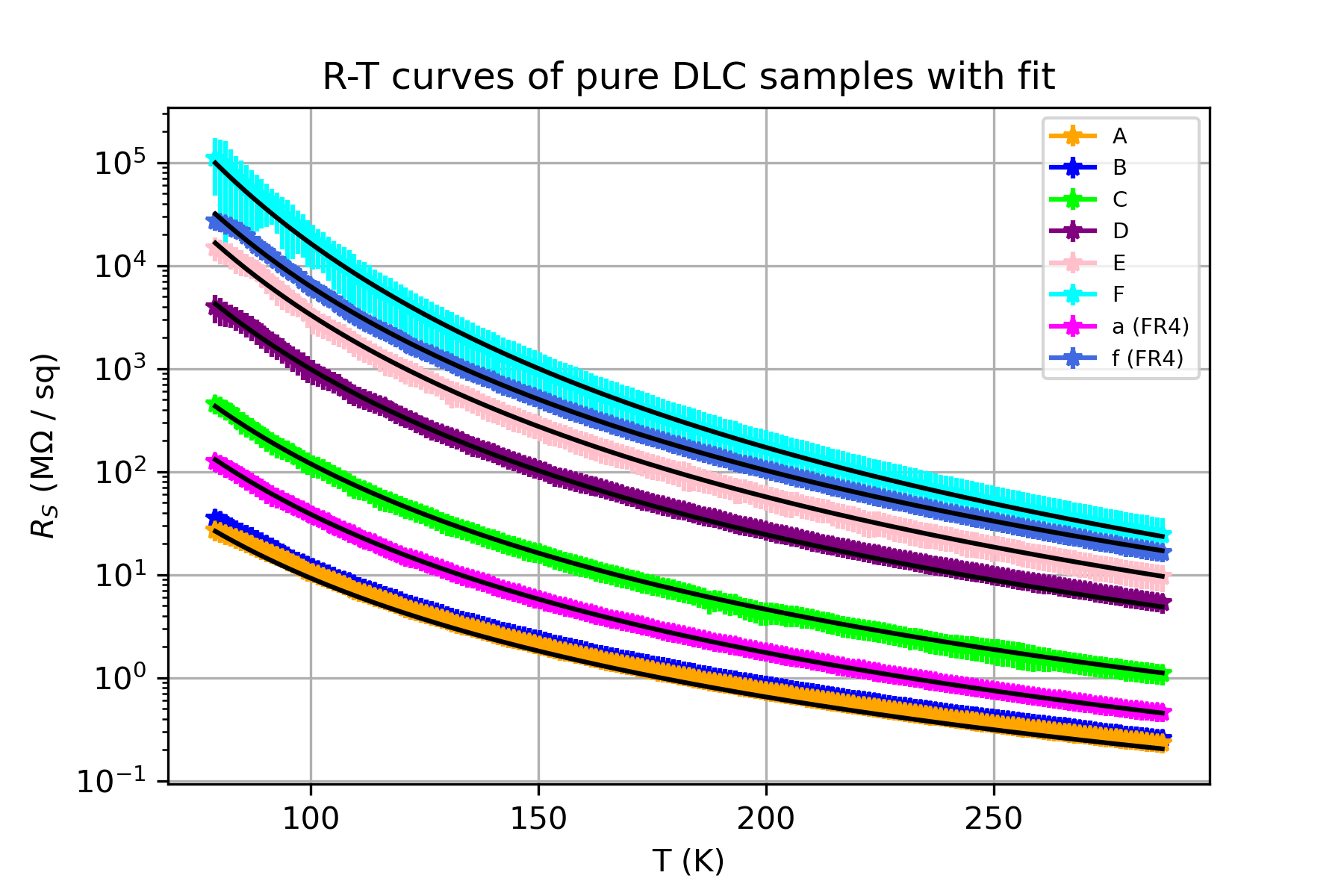}
    \centering
    \caption{\footnotesize $R$-$T$ curves for all DLC-coated samples, together with a fit to a 2-dimensional variable-range electron hopping model (black solid lines),
    following eq. \ref{eq:fitting}.}
    \label{fig:model1}
\end{figure}

$R$-$T$ curves are shown in Fig. \ref{fig:model1}, with error bars estimated from the deviation between two independent measurements taken with the setup described in Fig. \ref{fig:setup1} (for the A sample, the four available measurements from the PPMS were used instead).

For thin coatings based on amorphous carbon it has been shown, using for instance percolation theory, that the coating resistance is highly anisotropic with the following dependence:
\begin{equation}
    R = R_{300} * \exp{ \Bigg(\Big(\frac{T_0}{T}\Big)^a - \Big(\frac{T_0}{300}\Big)^a \Bigg)}
    \label{eq:fitting}
\end{equation}
being $a=1/4$ for the resistance across the bulk (3D) and $a=1/3$ for the resistance over the surface (2D) \cite{Book}. The form of eq. \ref{eq:fitting} has been chosen such that the pre-factor $R_{300}$ represents the surface resistance at 300~K (dubbed $R_{S,300}$ in the following). $T_0$ is an effective temperature given by:
\begin{equation}
T_0 = \frac{\beta}{k_B \cdot g \cdot r}
\end{equation}
where $k_B$ is the Boltzmann constant, $\beta$ a factor of order 1, $g$ the density of states at the Fermi level of the film material, and $r$ the localization radius around it. Following the original naming by Mott, this type of conduction mechanism is widely known as variable-range hopping (vrh), and we will consider it as the starting point for the interpretation of DLC data.

\begin{figure}[H]
    \centering
    \includegraphics[width=14cm]{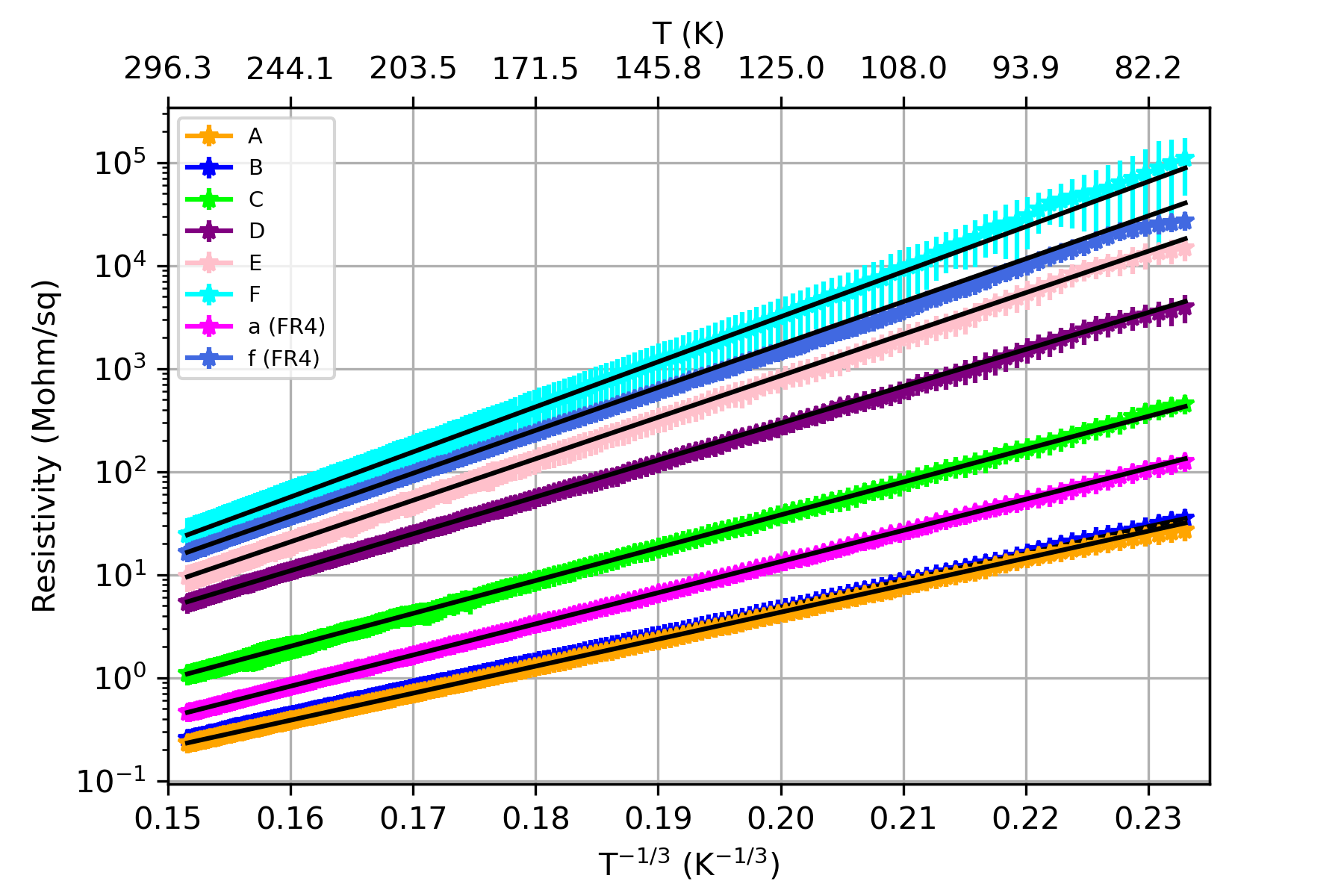}
    \centering
    \caption{\footnotesize $R_S$ as a function of $T^{-1/3}$ (lower x-axis) and $T$ (upper x-axis) in a logarithmic representation, demonstrating the 2-dimensional variable range hopping assumption. Only for very low temperatures small deviations appear, an effect likely attributable to the measurement setup.}
    \label{fig:model2}
\end{figure}

\begin{figure}[H]
    \centering
    \includegraphics[width=14cm]{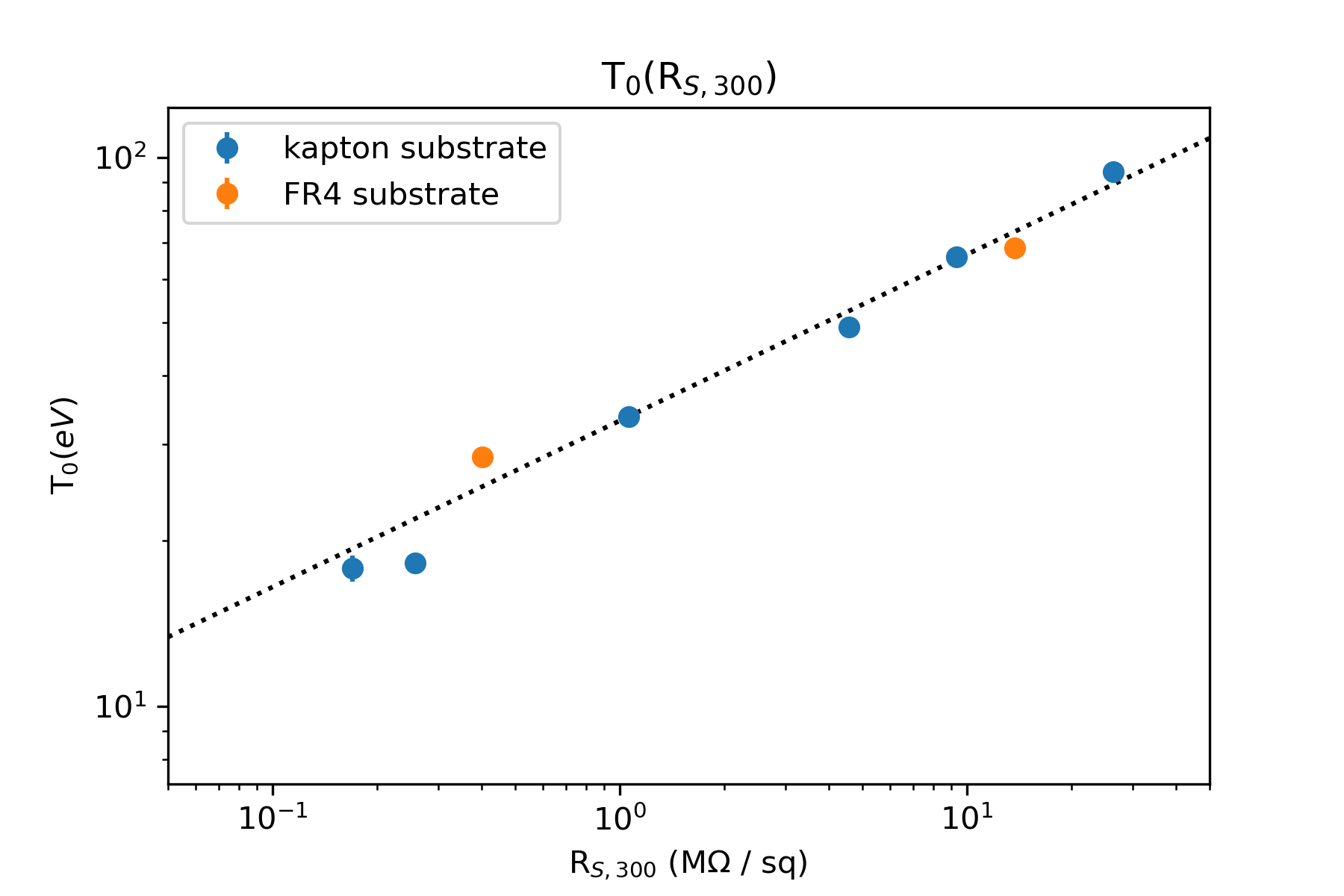}
    \centering
    \caption{\footnotesize Parameter $T_0$ as a function of the room temperature resistivity, $R_{S,300}$, together with a fit to a power-law. Blue points refer to the Kapton substrate and orange ones to FR4. }
    \label{fig:Tzeroofrho}
\end{figure}

A logarithmic representation of the resistivity curves as a function of $T^{-1/3}$ confirms in fact the expectation from 2D-vrh (Fig. \ref{fig:model2}), reproducing early observations by Hauser on amorphous carbon \cite{Hauser}. The parameters $T_0$ and $R_{S,300}$, found by fitting each curve individually and keeping $a = 1/3$, display a power-law correlation in our data, as $T_0 \sim R_{S,300}^\nu$, with $\nu=0.3$ (Fig. \ref{fig:Tzeroofrho}). Such a power-law dependence is expected in 2D-vrh conduction (e.g., eq. 9.4.5 in \cite{Book}).

\begin{figure}[H]
    \centering
    \includegraphics[width=14cm]{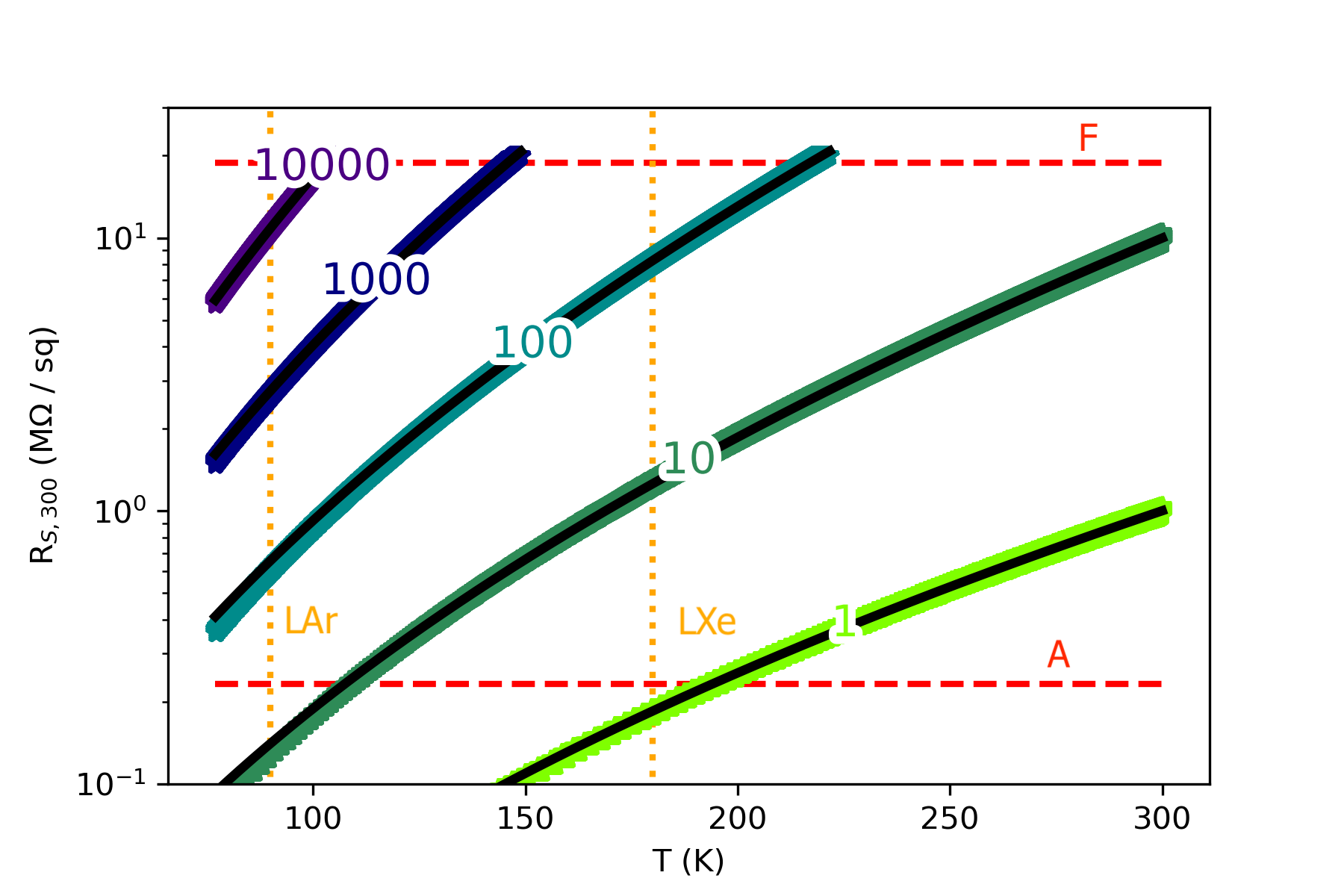}
    \centering
    \caption{\footnotesize Iso-resistivity curves as a function of temperature, generated numerically. Once the desired $R_S$-value  (1, 10, 100, 1000, 10000 ~M$\Omega/\Box$) is chosen for a certain operating temperature, the corresponding room temperature value ($R_{S,300}$) can be determined. The latter is in principle configurable at the sputtering machine, through the total sputtering time. Overlapped to the numerically-generated curves, the black line is the approximate fit to function \ref{eq:prediction}. Dashed lines $A$ and $F$ represent the two most extreme $R_{S,300}$ values in this study.}
    \label{fig:prediction}
\end{figure}

Using function \ref{eq:fitting} and the relation between $T_0$ and $R_{S,300}$, iso-resistivity curves can be obtained numerically for each starting value of $R_{S,300}$ (Fig. \ref{fig:prediction}). These numerically-generated curves are well described by a function of the form 
\begin{equation}
   R_{S,300} = C * \exp(T^b)
    \label{eq:prediction}
\end{equation}
with $0.38 < b < 0.5$ in the range considered. This plot allows to predict the room-temperature surface resistivity that needs be targeted at the sputtering machine, to obtain the desired resistivity at the temperature of operation. With the range of $R_{S,300}$ studied in this work (top and bottom lines in the figure, corresponding to films $A$ and $F$), it is clear that the ranges of potential interest as per table \ref{tab:data} are reasonably well covered, in particular when aiming at operation at liquid argon and liquid xenon conditions.

\subsection{Long-term stability \label{sec:aging}}

Long-term stability is a crucial aspect for applicability in particle-physics experiments, expected to run over time spans of years. First, we excluded any important effect of humidity, by studying sample $B$ directly in open air (R.H. around 40\%) and in helium atmosphere, with the two measurements agreeing within 0.5\%. By simply storing the samples in plastic bags, for over more than 1 year, all kept their conductivity within less than a factor $\times 2$ variation.

\begin{figure}[H]
    \centering
    \includegraphics[width=0.49\textwidth,]{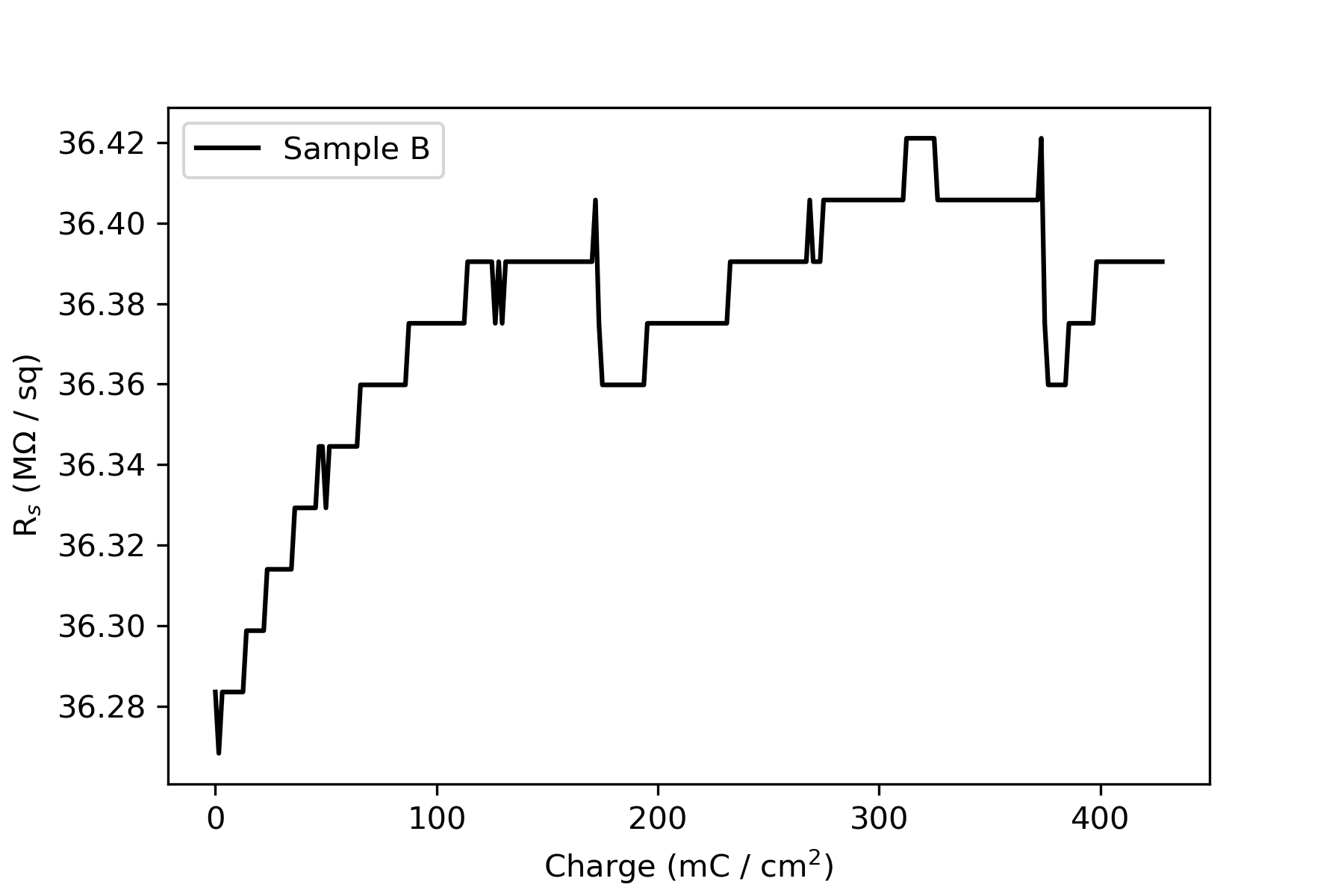}
    \includegraphics[width=0.49\textwidth]{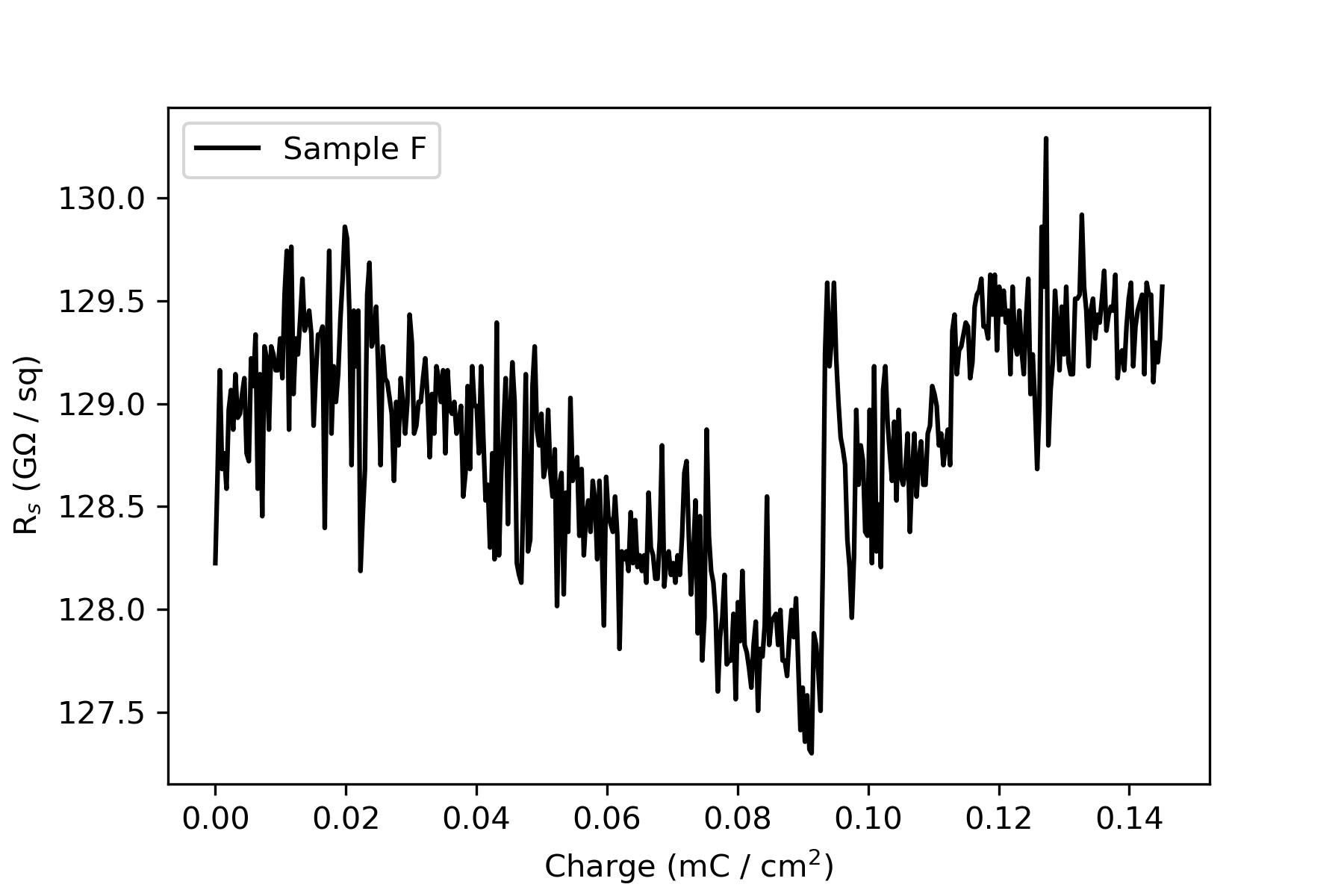}
  
    \caption{\footnotesize Long term stability test for samples $B$ (left) and $F$ (right) at LNT. Both samples show variations of current $\leq$ 2\% when 500 V are applied for more than 5 hours in a helium atmosphere.}
    \label{fig:aging}
\end{figure}

Another concern is the appearance of charge-depletion effects, that limit for instance ionic conductors. We studied for that the stability of one of the least and most resistive samples ($B$ and $F$) at $LN$ temperature, in the 2-stripe configuration (Fig. \ref{sec:setup}). To accelerate the migration of charge carriers, we applied 500~V for about five hours, while keeping the vessel inside $LN$. Both samples showed variations of current of $\leq$ 2\% (Fig. \ref{fig:aging}), which points towards a good stability of the material. Furthermore, measurements in helium atmosphere at room temperature, where the ageing can be greatly accelerated, 
showed no significant variation for transported charges as large as 2~C/cm$^2$. The absence of charge-depletion effects, combined with the observed $T$-dependence, are compatible with a 2-dimensional variable-range conduction mechanism based on electron hopping.

Although the transported charges in particular in Fig. \ref{fig:aging}-right are fairly low, it must be noted that cryogenic detectors often work at very low interaction rates. In the case of DUNE Far-Detector, for instance \cite{FD}, where background rates are expected to be dominated by $^{39}$Ar decays \cite{Jose}, the charge density at the anode of a resistive-protected multiplication structure (e.g., \cite{Roy}), even if assuming a (probably optimistic) gain of 100, would be about 60.5 $\mu$C/cm$^2$ over 10~years, surpassed in our ageing tests even for the most resistive film (Fig. \ref{fig:aging}-right). The calculation comes from a $^{39}$Ar decay rate of around 1.4 kHz/m$^3$ and an average energy per decay of 220 keV, assuming a 10~m drift \cite{BenettiGS}.

It must be noted, though, that given that the conduction mechanism of DLC does not seem to change in the $T$-range studied (as inferred from the good agreement with the 2D-vrh hypothesis), the stability observed up to 2~C/cm$^2$ at room temperature is expected to apply to $LN$ too, making these coatings in principle viable also in high-rate applications, at either temperature. Ageing effects related to chemical reactions or deposits from the avalanche process cannot be evaluated by this simple procedure, and require sustained detector operation.

\section{Results from a resistive-protected detector \label{sec:summary}}

Resistive-protected Micro-pattern Gas Detectors (MPGDs), for instance based on thick-GEM (THGEM) \cite{Bressler_2013}, GEM \cite{Bencivenni_2015} or Micromegas \cite{ResistiveMicromegas} technology, have been extensively studied at room temperature and their properties characterized. So far, the main limitation of exporting this technology to dual-phase liquid argon detectors (T=87.5~K at 1~bar) was the lack of suitable resistive materials (films, plates) with the right value of $R_S, \rho_v$. 
As discussed, DLC could in principle act as a spark-quenching element in such conditions.

\begin{figure}[H]
    \centering
    \includegraphics[width=10cm]{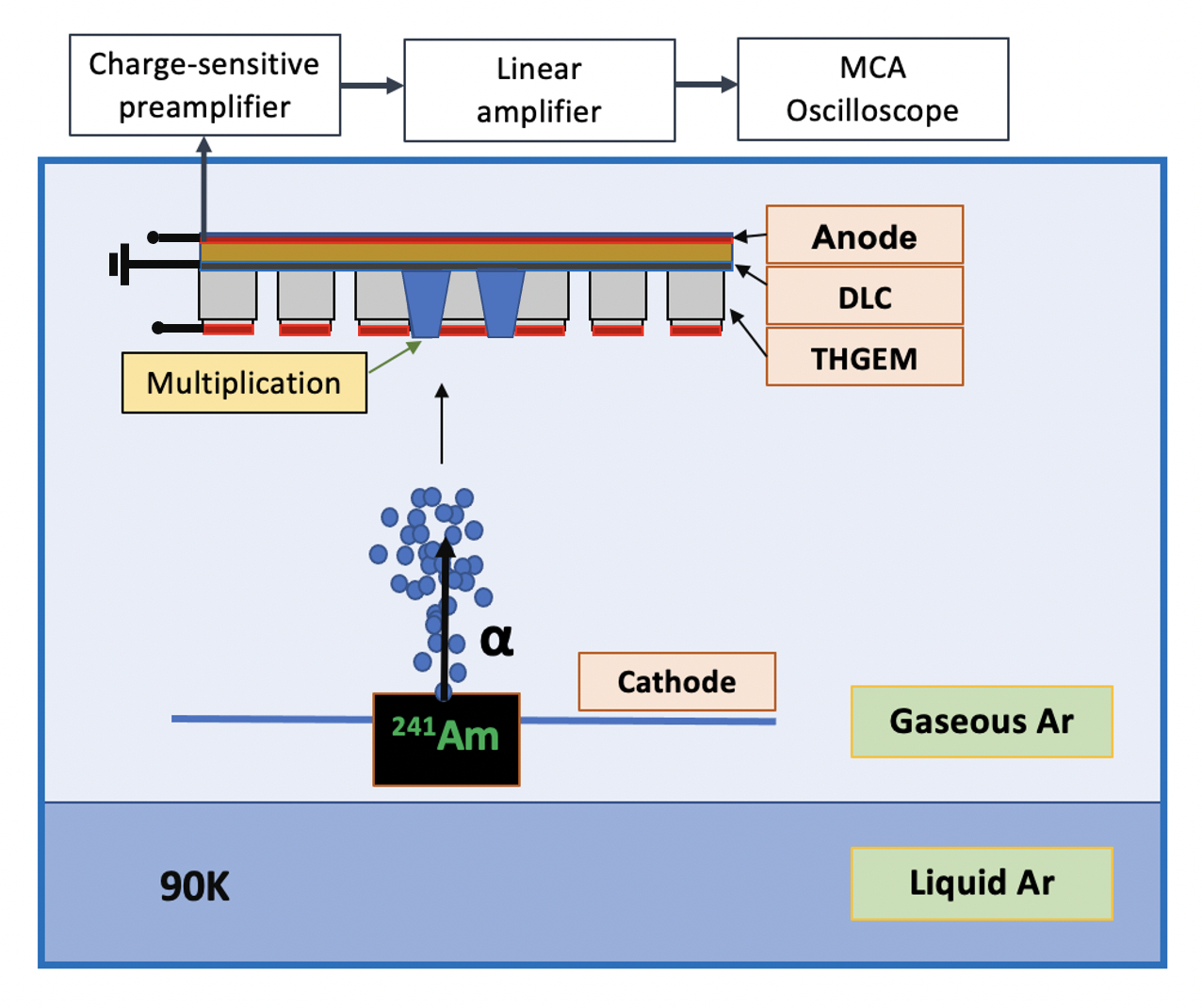}
    \centering
    \caption{\footnotesize Scheme of an RWELL detector under test in dual-phase argon.}
    \label{fig:RWELL}
\end{figure}

The resistive WELL (RWELL) \cite{Arazi_2014} was chosen for our first investigations. A 1~mm-thick THGEM electrode was coupled to a plate anode via a DLC film ($R_S \sim $20~MOhm/$\Box$ at 90~K, sample A), the latter deposited onto a 0.05 mm-thick Kapton foil fixed on a 1-mm thick insulating FR4 substrate.\footnote{As illustrated in Fig. \ref{fig:SEM}, the surface quality in case of direct DLC evaporation onto FR4 plates does not seem adequate for this particular application}
A $^{241}$Am source was installed on a metallic plate (cathode) 15~mm below the amplification structure and a field of 1~kV/cm was implemented in the drift region. A conceptual scheme of the cryogenic RWELL detector and test setup is depicted in figure \ref{fig:RWELL}.

\begin{figure}[H]
    \centering
   \includegraphics[width=10cm]{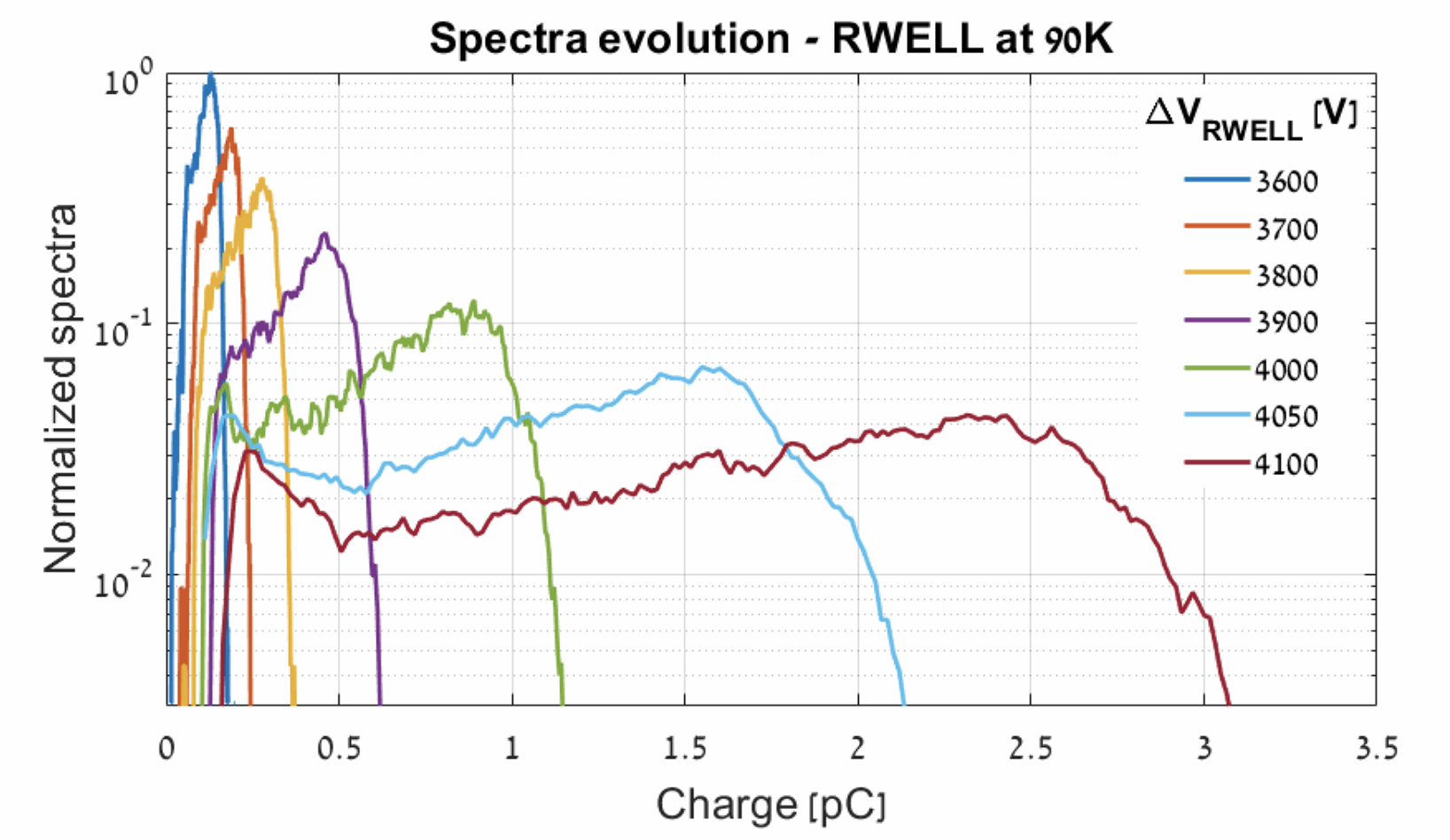}
    \centering
    \caption{\footnotesize Spectra evolution for different RWELL voltage configurations, under $\alpha$-events from $^{241}$Am.}
    \label{fig:alphaSpect}
\end{figure}

Ionization electrons released by $\alpha$-particles in the drift gap are collected into the RWELL holes: they undergo charge multiplication according to the voltage applied across the structure, and the signals induced can be recorded from the anode and processed with appropriate electronics. The high resistivity of the DLC film ensures a priori good signal transparency through it and occasional spark-quenching. A set of charge-spectra, containing 5000 waveforms each and corresponding to different RWELL voltage configurations, is shown in Fig. \ref{fig:alphaSpect}. Each spectrum is normalized to its area and to the maximum at $\Delta V_{RWELL}$ = 3600~V.

It is possible to notice that each $\alpha$-spectrum possesses a full-energy peak accompanied by a low-energy tail accounting for partial energy deposition; it is due to the isotropic emission of the source combined with the effect of the collimator. The spectra shifts to higher values with the applied voltage, as expected. Further details on the RWELL detector operation and stability at cryogenic temperatures will be provided in a dedicated publication.

\section{Conclusions}

We propose the use of resistive diamond-like carbon (DLC) coatings for applications related to gaseous and liquid particle detectors operating in cryogenic conditions, down to liquid nitrogen temperature. Our characterization of DLC samples on Kapton substrates (10 $\times$ 10 cm$^2$) reveals a good behaviour in terms of linearity, surface uniformity (electrical and physical), easiness of adjustment of the electrical properties, and long-term stability across a range of temperatures; these add to other properties well established in literature such as low chemical reactivity and radiation tolerance. In particular, the surface-resistivity range $R_S = 1$-$10000$~MOhm/$\Box$ has been reasonably covered by the films studied here, in the temperature range comprising the LXe and LAr liquefaction points. This resistivity range is expected to cover a variety of problems beyond spark-quenching: examples are cluster-size tuning, field-cage or feed-through lining, and more. 

The electrical behaviour with temperature and time is compatible with a conductivity mechanism based on 2-dimensional variable-range electron hopping. This important observation opens the path to a better and more detailed understanding of the surface and bulk conductivities of DLC coatings used in modern particle-physics instrumentation.

First results from a DLC-equipped RWELL based on this technology, close to the liquid-vapour equilibrium point of argon, have been presented.

\section{Acknowledgments}

We are thankful to the superconductivity group of USC (in particular to J. Mosqueira) for encouragement, discussions, and hardware support. We thank the skillful work of Bruno Da Cuña and Ramiro Barreiro on sample analysis as well as the availability of analytical facilities (such as the PPMS system) of USC-RIAIDT-USC. Enlightening conversations with J. Maneira (LIP) on the expected DUNE rates were essential to put the long-term study in context. This research has been sponsored by RD51 funds through its `common project' initiative, and has received financial support from Xunta de Galicia (Centro singular de investigación de Galicia accreditation 2019-2022), and by the “María de Maeztu” Units of Excellence program MDM-2016-0692. DGD is supported by the Ram\'on y Cajal program (Spain) under contract number RYC-2015-18820.

\bibliographystyle{elsarticle-num}

\end{document}